\documentclass{iopart}
\usepackage{iopams}
 \pdfoutput=1  
\usepackage{amssymb}
\usepackage{graphicx}
\usepackage{cite}

\newcommand{\textnew}[1]{\mathrm{#1}}
\newcommand{\mphi}{\phi}
\newcommand{\Sbraid}{S_{\textnew{braid}}}
\newcommand{\logLt}{\log L(t)}

\newcommand{\MSD}{\sigma^2}

\usepackage[usenames,dvipsnames]{color}
\newcommand{\coloronline}{(Color online)\quad}

\begin{document}

\title{Trajectory entanglement in dense granular materials}

\author{James G. Puckett$^1$, Fr\'ed\'eric Lechenault$^2$, Karen E. Daniels$^1$, Jean-Luc Thiffeault$^3$}%
\address{$^1$Department of Physics, NC State University, Raleigh, NC, 27695 USA }
\address{$^2$Laboratoires de Physique Statistique, \'Ecole Normale Sup\'erieure, 24 rue Lhomond, 75005 Paris, France}
\address{$^3$Department of Mathematics, University of Wisconsin, Madison, WI, 53706 USA }

\begin{abstract}
  The particle-scale dynamics of granular materials have commonly been
  characterized by the self-diffusion coefficient $D$. However, this
  measure discards the collective and topological information known to
  be an important characteristic of particle trajectories in dense
  systems. Direct measurement of the entanglement of particle
  space-time trajectories can be obtained via the topological braid
  entropy $\Sbraid$, which has previously been used to quantify mixing
  efficiency in fluid systems. Here, we investigate the utility of
  $\Sbraid$ in characterizing the dynamics of a dense, driven granular
  material at packing densities near the static jamming point
  $\phi_J$. From particle trajectories measured within a
  two-dimensional granular material, we typically observe that
  $\Sbraid$ is well-defined and extensive.
However, for systems
  where $\phi \gtrsim 0.79$, we find that $\Sbraid$ 
(like $D$) is not well-defined,
  signifying that these systems are not ergodic on the experimental
  timescale. Both $\Sbraid$ and $D$ decrease with either increasing
  packing density or confining pressure, independent of the applied
  boundary condition.  The related braiding factor provides a
  means to identify multi-particle phenomena such as collective
  rearrangements.  We discuss possible uses for this measure in
  characterizing granular systems.
\end{abstract}




\maketitle

\section{Introduction}

Recently, there have been extensive efforts to draw parallels between dense granular systems, foams, emulsions, and glassy molecular systems \cite{Liu2010,vanHecke2010}. For such systems approaching the glass/jamming
transition, the dynamics slow down due to an increasing intrication of
the available phase space, with rearrangements taking the form of cage jumps \cite{Weeks2000, Weeks2002, Pouliquen2003, VollmayrLee2004, Marty2005, Reis2007a}.  In this regime, particles are trapped by their neighbors at short timescales and resume diffusive behavior at long timescales due to eventual loss of correlation in the succession of jumps.  Such jumps involve cooperatively-rearranging regions which induce dynamical heterogeneities \cite{Cipelletti2005, Chandler2010}, and the particles pass each other only rarely. The growth in size of these rearranging regions is believed to be associated with the rigidity and loss of ergodicity exhibited by fragile glasses.

Unlike glassy molecular systems, granular materials permit the tracking of individual particles. The average dynamics have traditionally been characterized by the mean squared displacement, $\MSD(\tau)$.  When the experimental timescale is long enough compared to the viscous timescale, permitting a complete decorrelation in the particle motion, it is possible to calculate the self-diffusion constant $D$.  This coefficient is a \emph{single}-particle measure in which all information regarding relative motions of particles is discarded.                                                                                                                                                                                                                                                                                                                                                                                                                                                                                                                                                                     
In contrast, recent experiments and simulations \cite{Weeks2000, Reis2007a, Berthier2005, Dauchot2005, Keys2007, Lechenault2008, Candelier2009a, Candelier2010a, Duri2009, Candelier2010}
have shown that the dynamics near the glass transition become heterogeneous in both space and time. As such, it is necessary to take relative motions into account when formulating the correct average description of the state. To this end, \emph{multi}-particle scalars have been developed, including the four-point susceptibility $\chi_4$ \cite{Dauchot2005, Keys2007, Lechenault2008}. However, to our knowledge, no scalars have been introduced which characterize mixing effects, which may be relevant to quantifying the cooperative motion of particles.

The central observation of this paper is that the trajectories of a two-dimensional assembly of grains take the form of a \emph{braid} when plotted in a space-time diagram, as shown in Figure~\ref{fig:apparatus}. Each strand of the braid represents the space-time trajectory of a single particle in the system, with the length of the braid corresponding to time and the entanglement of the strands arising from dynamics. The topological braid entropy, $\Sbraid$, is a measure of the degree of this entanglement. This approach has been successfully used to analyze the mixing efficiency in two-dimensional fluid flows \cite{Boyland2000, Thiffeault2005, Gouillart2006, Thiffeault2006, Thiffeault2010, Allshouse2011}. Specifically, the degree of lagrangian chaos created by the motion of rods in a fluid is quantitatively related to the topology of the corresponding braid. We anticipate that this notion might also be applicable to characterizing caging and dynamical heterogeneity in granular systems, by providing a quantitative measure of how close a system is to departing from ergodicity.  

Here, we experimentally investigate the behavior of the average
topological braid entropy extracted from the long-time dynamics of
particles in a dense, driven granular material, examining the
dependence on boundary conditions (constant pressure, CP, and constant
volume, CV) and the inter-particle friction coefficient $\mu$.  We
analyze the dynamics with two methods: a single-particle method (the
self-diffusion constant, $D$) and a \emph{topological} multi-particle
method (the braid entropy, $\Sbraid$).  We find that both $D$ and
$\Sbraid$ are sensitive to the packing density $\mphi$ and pressure
$P$ of the system in the same qualitative way. However, at high
$\mphi$ and $P$, both $D$ and $\Sbraid$ are inaccessible due to
insufficiently long experimental timescales or
ill-definedness. In contrast, the {\it braiding factor} can be
readily computed and offers an instantaneous view of the magnitude and
intermittency of the rearrangements in the system.

\section{Experiment}

\begin{figure}[b]
\begin{minipage}[b]{0.65\linewidth}
\centering
\includegraphics[width=0.95\linewidth]{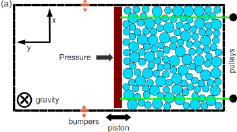}
\includegraphics[width=0.95\linewidth]{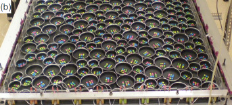}
\end{minipage}
\begin{minipage}[b]{0.32\linewidth}
\includegraphics[width=0.9\linewidth]{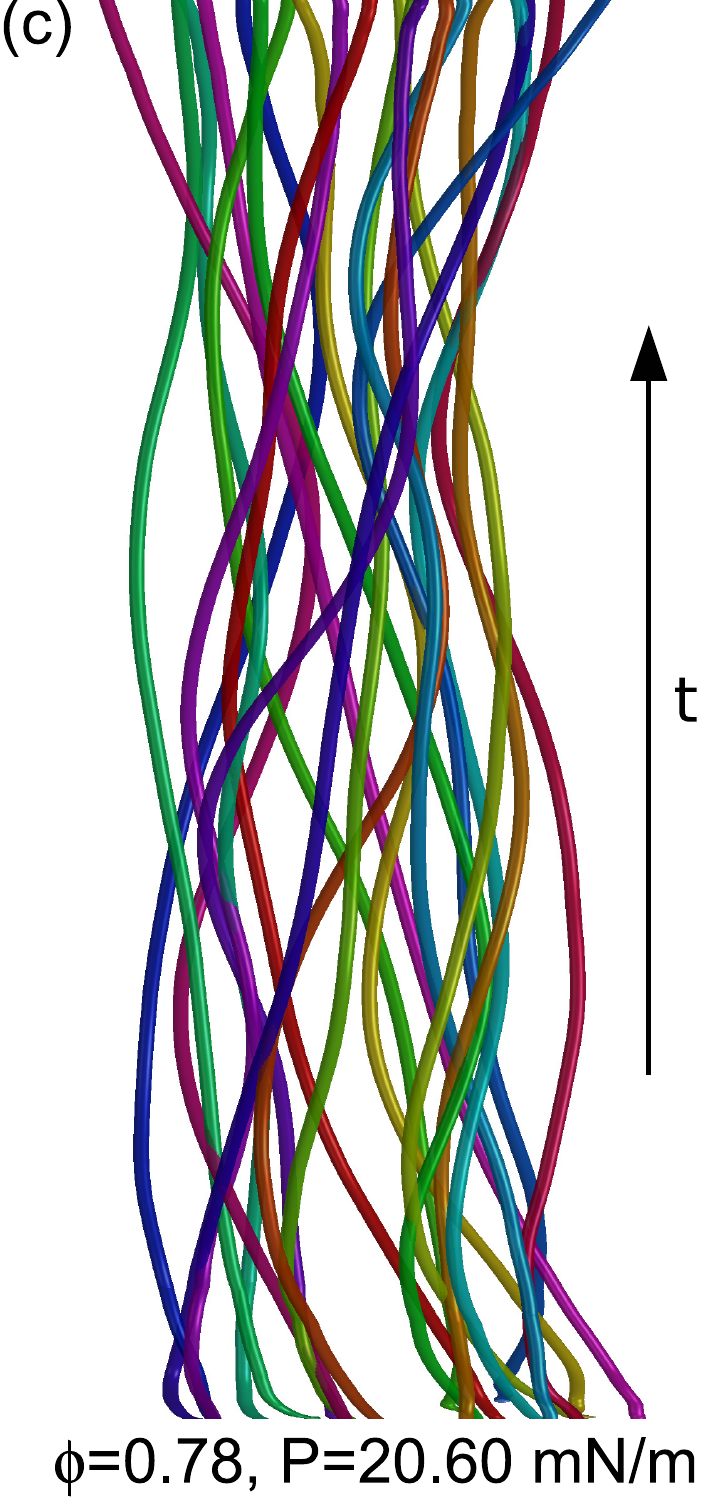}
\end{minipage}
\caption{\coloronline (a) Schematic and (b) photograph of apparatus. The piston provides constant pressure via pulleys and weights, or constant volume by fixing its position to the surface of the table. (c) Entanglement of space-time trajectories for $N=20$ particles.}
\label{fig:apparatus}
\end{figure}

We conduct experiments on a two-dimensional granular assembly supported upon a nearly-frictionless horizontal air table and agitated by bumpers around the boundary.  The granular material is bi-disperse, consisting of large and small particles with diameter of $r_L = 83.6$\,mm and $r_S = 55.8$\,mm and mass of $m_L = 8.1$\,g and $m_S = 3.5$\,g, respectively. The concentration of small/large grains is fixed at $N_S/N_L = 2$, so as to occupy similar areas. The apparatus is shown in Figure~\ref{fig:apparatus}.

The boundary condition and the inter-particle friction $\mu$ can be changed for each experimental run.  The boundary condition is established by the configuration of the piston ($m_{\textnew{piston}} = 95\ m_S$).   By attaching weights to the piston via a low-friction pulley, the system is confined under constant pressure (CP) with a range of pressures $5.7 < P < 80$\,mN/m. The constant volume (CV) boundary condition is established by fixing the piston to the table, where a range of packing densities $\mphi = 0.72$ to $0.81$ is explored by removing units of particles (two small and one large), maintaining the relative concentration. The inter-particle friction is selected by changing the material around the particles' outer edge, with $\mu_1 = 0.1$ for PTFE (Teflon) wrapping, $\mu_2 = 0.5$ for bare polystyrene, and $\mu_3 = 0.85$ for rubber.  

Around the boundary, an array of bumpers agitates the system on three sides.  Bumpers are triggered pairwise via a computer at a high frequency ($f = 10$ Hz) to keep particles in motion despite energy dissipated in collisions.  A triggered pair consists of a bumper and its corresponding bumper on the opposite wall. We have previously measured \cite{Nichol2011} that the effect of this driving system is to provide a thermal-like bath which maintains the granular material at constant average kinetic energy. More information on the driving and kinetics of the system can be found in references \cite{Nichol2011,Lechenault2010, Puckett2011}.

Images are taken $\tau_{\textnew{camera}} = 2$ or $5$ seconds apart depending on the $\mphi$-dependent dynamical timescales \cite{Lechenault2010}. We record the positions of particles and map trajectories in time using unique particle identifiers \cite{Lechenault2010}.  The unique identifiers allow a relatively large $\tau_{\textnew{camera}}$, so we can confidently track the trajectories of individual particles and thereby be certain that topological dynamics do not erroneously result from the exchange of particle identities due to tracking mistakes.  The packing density $\mphi$ is calculated as $\mphi=\langle \pi r_i^2 / V_i \rangle_i$ with $r_i$ the particle radius and $V_i$ its local Voronoi cell area, calculated with Voro++~\cite{Rycroft2009}.  The average is taken over all particles in the central $20 \%$ area of the table to reduce ordering effects induced by the boundaries \cite{Nichol2011, Desmond2009}.

\section{Results}

\subsection{Self-diffusion}

\begin{figure}
\includegraphics[width=0.95\linewidth]{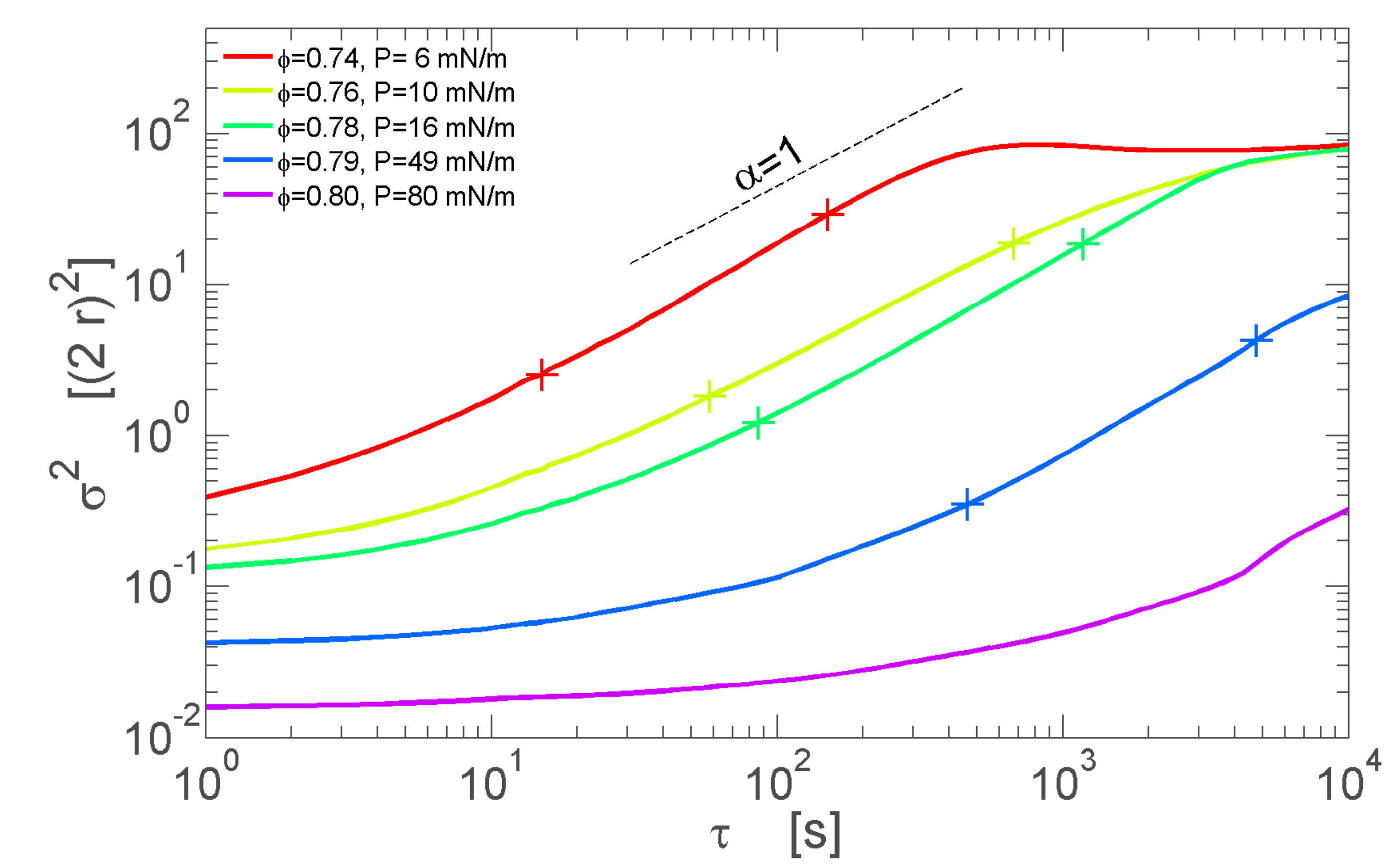}
\caption{Representative diffusion relations $\MSD(\tau)$ for constant pressure (CP) experiments with $\mu_2$ and $\mphi = 0.74$ to $0.80$. The dashed line represents a slope of unity. The diffusion coefficient, $D$, is calculated between the two $+$ symbols on each curve. 
}
\label{fig:diffusion}
\end{figure}

The diffusion behavior is quantified using the mean-squared displacement
\begin{equation}
\MSD(\tau)=\langle \Delta x_i(t_j+\tau)^2+ \Delta y_i(t_j+\tau)^2 \rangle_{i,j}
\label{eqn:msd}
\end{equation}
where the average $\langle \cdot \rangle$ is performed over all particles $i$ and all times $t_j$.  As can be seen in Figure~\ref{fig:diffusion}, the slope of $\MSD(\tau)$ is not constant for all timescales $\tau$. It is useful to define a local curve $\MSD(\tau) \propto \tau^\alpha$, where the value of $\alpha$ characterizes the average individual dynamics associated with that $\tau$. For short timescales ($\tau < \tau_D$), the trajectories remain sub-diffusive, with $\alpha < 1$.  This arises because the particles are locally confined, or caged, by neighboring particles \cite{Weeks2000, Weeks2002}. At longer timescales ($\tau > \tau_D$), each particle has experienced multiple cage-breaking events and thus resumes diffusive behavior. For timescales at which $\alpha = 1$, we can measure the two-dimensional self-diffusion coefficient $D = \frac{1}{4}\frac{d}{d\tau} \MSD(\tau)$. Finally, if the diffusion process is fast enough, a typical particle covers a distance of many particle diameters within the duration of the experiment.  In such cases, the exponent $\alpha$ can again fall below unity at long times, due to the finite size of the experiment. This effect is visible in Figure~\ref{fig:diffusion} for the lowest $\mphi$ at which we perform experiments.

In general, it is possible to calculate $D$ using $\MSD(\tau)$ for experiments which are of long enough duration to reach the $\alpha = 1$ regime. For this system, we find that this regime is reached even at values of $\mphi$ near static random loose packing \cite{Lechenault2010}. A convenient means to measure $D$ is to specify a tolerance on $\alpha$ which defines a diffusive regime, marked by $+$ symbols in Figure~\ref{fig:diffusion}. We calculate $D$ by fitting to the largest continuous region where the smoothed derivative of $\MSD(\tau)$ has a slope $| \alpha - 1  | < 0.1$. The diffusive timescale $\tau_D$ is the timescale at which the particle motion first achieves diffusive behavior.  With increasing packing density $\mphi$, $\tau_D$ becomes longer, as seen in Figure~\ref{fig:diffusion}.  For systems with large $\mphi$ and $P$, the range of $\tau$ over which $\MSD(\tau)$ was used to find $D$ (marked with crosses in Figure~\ref{fig:diffusion}) becomes less than a decade.  

Given the constraint that $| \alpha-1 | < 0.1$ and the finite duration of our experiment $\approx 10^4$\,s, it is not always possible to calculate a diffusion constant for systems with large $P$ or $\mphi$. One such example ($\mphi = 0.80$, $P = 80$\,mN/m) is shown in Figure~\ref{fig:diffusion}.  A more conservative tolerance specification on $\alpha$ would further decrease the experiments for which we could calculate a meaningful value of $D$.  In many experiments where we would wish to quantify the self-diffusion of particles, $D$ is in fact poorly-defined.

\subsection{Braid entropy}

The braid entropy is a measure of the \emph{degree of entanglement} of trajectories \cite{Thiffeault2005,Thiffeault2006,Thiffeault2010}.  To define it, first consider an imaginary rubber band enclosing the strings at the bottom of the braid shown in Figure~\ref{fig:apparatus}c.  As we slide the rubber band upwards, its
length~$L(t)$ will tend to grow, as it is caught on the strings and is
not allowed to traverse them.  If the trajectories are chaotic,
then~$L(t)$ will grow exponentially; the growth rate is then related to the Lyapunov exponent of a dynamical system. The braid entropy~$\Sbraid$ is defined as the asymptotic growth rate of~$L(t)$,
\begin{equation}
\Sbraid = \lim_{t\rightarrow\infty}\frac{d}{d t} \logLt,
\label{eq:Sbraid}
\end{equation}
maximized over the choice of rubber band. This rubber band viewpoint
is well-illustrated by taffy pullers and dough-kneading devices, where
material is made to stretch exponentially by repeated folding \cite{Finn2011}.  The length
$L(t)$ is called the \emph{braiding factor}.

When trajectories arise from a continuous dynamical system, then the
braid entropy is a lower bound on the \emph{topological entropy} of
the system.  The topological entropy is a measure of the loss of
information about the identity of trajectories --- it is an entropy in
the sense of information theory.  As more trajectories are included,
the braid entropy converges to the topological entropy~\cite{Cox2007}.
However, in a granular medium such as the one considered here, there is
no underlying continuous dynamical system: the braid entropy is
not necessarily related to some intrinsic topological entropy.

\begin{figure}
\includegraphics[width=0.95\linewidth]{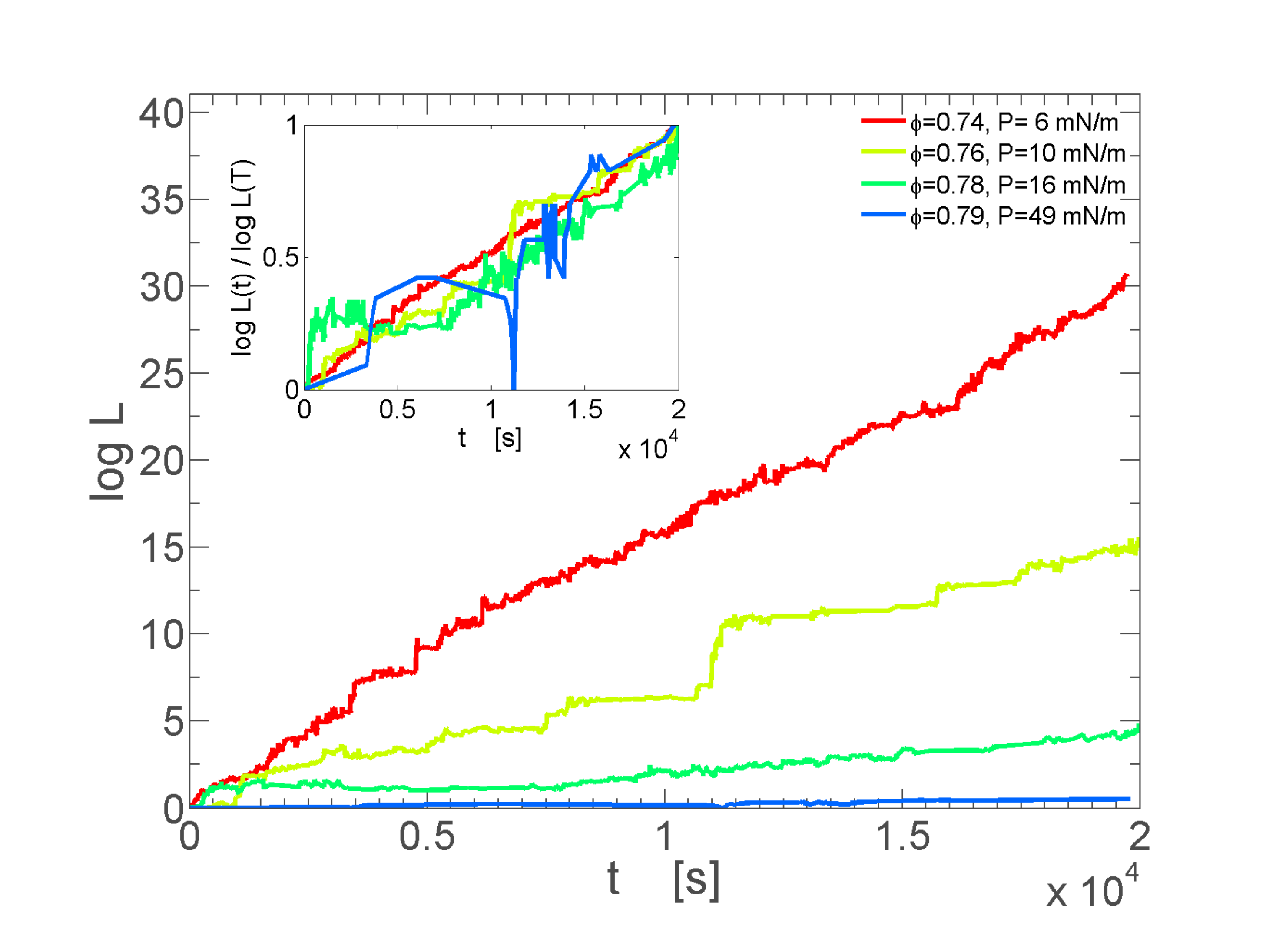}
\caption{The braiding factor $\logLt$ for the same experimental runs shown in Figure~\ref{fig:diffusion}, with $\mphi = 0.74$ to $0.79$.  The inset shows $\logLt$ divided by its value at the maximum time, $T$.}
\label{fig:braid}
\end{figure}

We follow the method described in \cite{Thiffeault2010} to calculate a
good approximation to the braid entropy.  This is a symbolic approach
based on earlier work~\cite{Dynnikov2002,Moussafir2006} where a braid
is first converted into its algebraic group representation for a particular
choice of projection axis, and then used to measure the growth of
hypothetical loops similar to the rubber band described above. We
calculate the braiding factor $L(t)$ using the trajectories of $N =
20$ particles. Because particles can enter and leave the region
captured by the camera, this choice of number provides a compromise
between having too many particles with missing points in the
trajectory, and too few particles with the maximum trajectory
duration. We find that this choice of $N$ is sufficient for $L(t)$ to
capture the dynamics of the whole system, as will be discussed in more
detail below.

The algebraic group representation is constructed from a list of the
crossing times and identities of the particles involved in each
crossing. Linear interpolation is used to determine the
instant of crossing along the projection line, and thus obtain the
braid generators. The experimental data is sufficiently well-resolved
for this method \cite{Thiffeault2010} to accurately capture crossings.

The braid entropy $\Sbraid$ (see Equation~\ref{eq:Sbraid}) is measured from the growth rate of the logarithm of the braiding factor. To obtain a single value for a particular run, we calculate a linear fit to $\logLt$. This entropy will be well-defined  if $\logLt$ meets three conditions: grows linearly in time, is an extensive quantity, and is independent of the choice of projection axis. We can directly check each of these conditions.

In Figure~\ref{fig:braid}, we show $\logLt$ for systems identical to Figure~\ref{fig:diffusion}, omitting the ($\mphi = 0.80$, $P = 80\ $mN/m) system as  too few persistent crossings occur.  We find $\logLt$ grows, on average, linearly with time for $\mphi \lesssim 0.79$. In the inset, $\logLt$ is divided by its value at the maximum time, to highlights how the degree of linearity present at different values of $\phi$.  

\begin{figure}
\center{\includegraphics[width=0.75\linewidth]{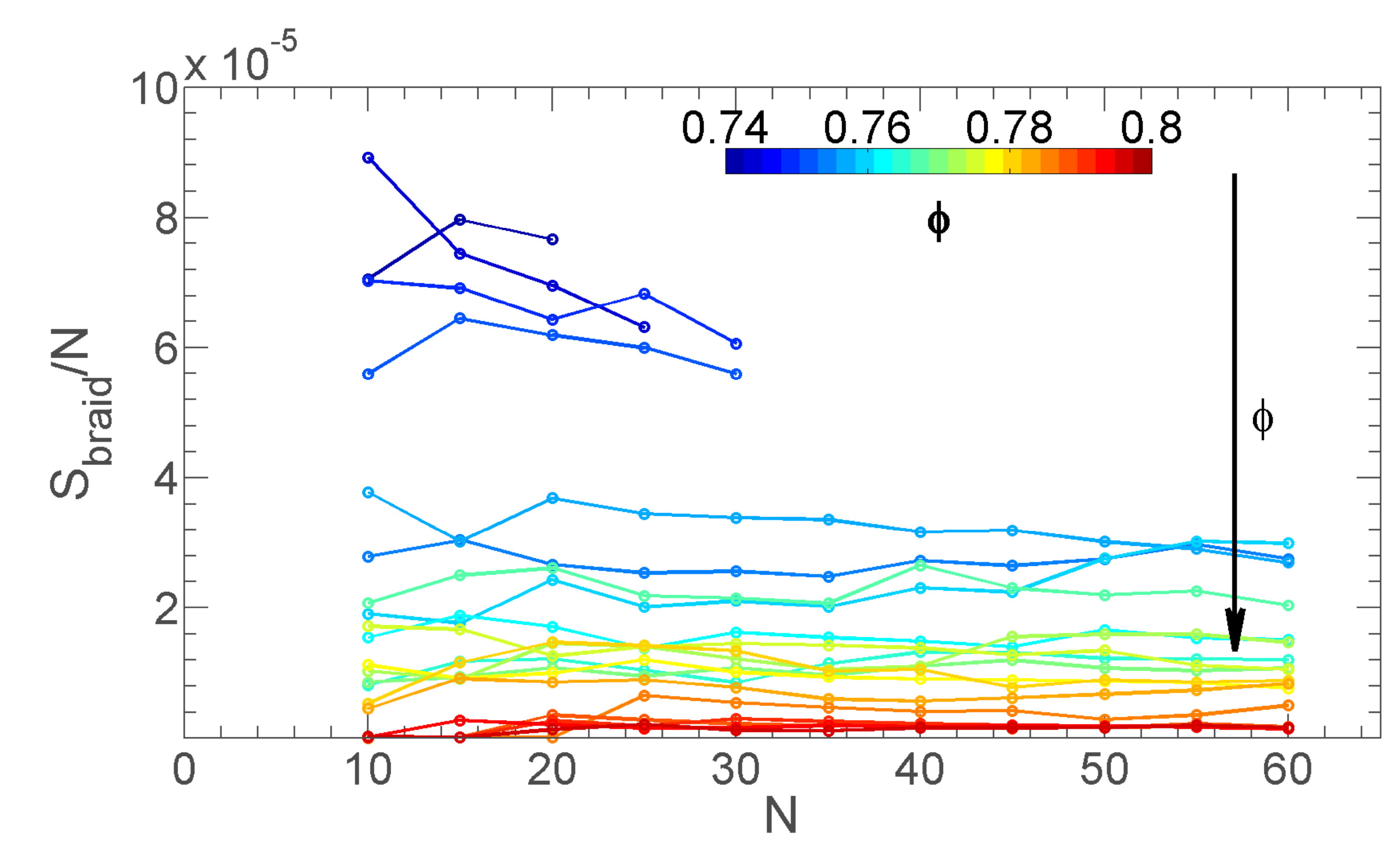}}
\caption{The $\Sbraid$ per particle, shown for $\mu_2$ and CP systems, with the axis of projection $\theta=0$. The arrow shows increasing $\mphi$ of the system.}
\label{fig:sovern}
\end{figure}

To test the extensivity of $\Sbraid$, we repeat the calculation described above for different numbers of particle trajectories, and test whether $S/N$ is a constant. Figure~\ref{fig:sovern} shows the results of this test for a series of CP experiments at different $\mphi$. For $N \gtrsim  15$, we observe that $S/N$ is quite flat, independent of boundary condition, $\mu$, $\mphi$, and $P$. This observation is consistent with $\Sbraid$ being an extensive quantity. For very small ensembles of trajectories, $N \lesssim 15$ (less than 10\% of the system size),  we find that the braid is too sparse to be representative of the dynamics of the system. In addition, it is difficult to test for extensivity in systems with low $\mphi$. In this regime, particles have a larger diffusion constant and are therefore more likely to enter and leave the region being monitored by the camera. Therefore, it is not possible to find $N \gtrsim 25$ particles which all satisfy the constraint on the minimum duration of continuous trajectories required to make a calculation of $\Sbraid$.

\begin{figure}
\parbox{0.5\textwidth}{\includegraphics[width=0.95\linewidth]{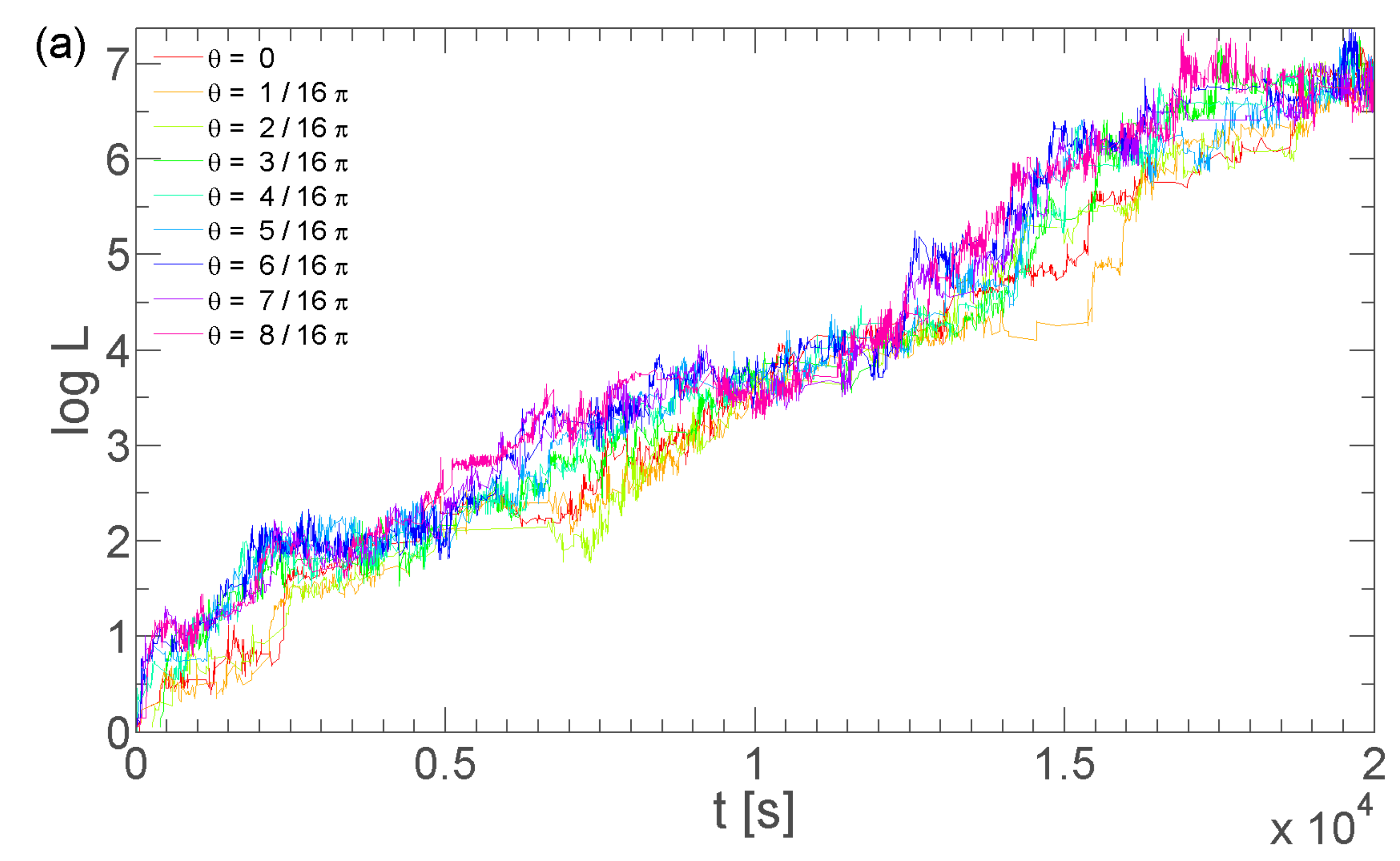}}
\parbox{0.5\textwidth}{\includegraphics[width=0.95\linewidth]{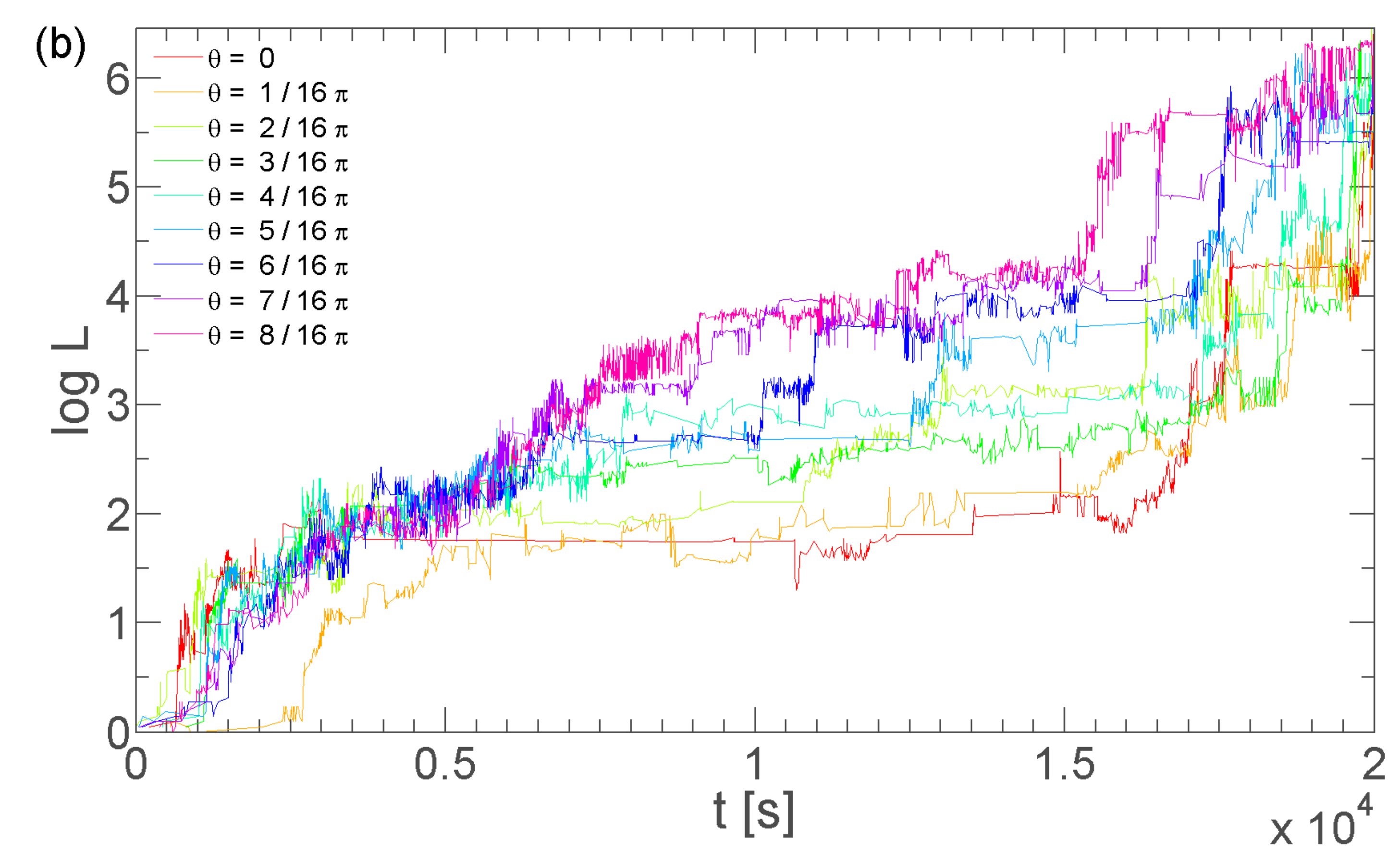}}
\caption{A plot of $\logLt$ for two systems with $\mu_2$ where the axis of projection is rotated between $\theta=[0,\pi/2]$, where (a) is $\mphi=0.77, P= 12$~mN/m and (b) is $\mphi=0.78, P= 18$~mN/m. 
}
\label{fig:srotation}
\end{figure}

We compute $L(t)$ for $9$ projections along axes oriented from $\theta = 0$ to $\pi/2$, and  compare the results in Figure~\ref{fig:srotation} for two characteristic systems, both confined at CP and with $\mu_2$ but at different $\mphi$ and $P$. For the example at low $\mphi$, we observe that $\logLt$ is quite linear, even on short timescales, and that the choice of projection axis has little effect on the slope of $\logLt$ or $\Sbraid$. However, in the higher-$\mphi$ example, $\logLt$ grows more quickly for the $(\theta=\pi/2)$-projection than for the $(\theta=0)$-projection, where the difference in entropy is $\approx 2\%$.  Increasing $N$ decreases the difference in entropy between projections.  In the limit of long experiment duration, such differences in calculated entropies would be expected to vanish.

In examining the runs at  large $\mphi$ and large $P$, we note the occurrence of two important features in $\logLt$: flat regions and steep increases. In Figure~\ref{fig:srotation}b (representative of such systems), we observe that $\logLt$ may remain constant for an extended period of  time, as in this case for  the interval $0.7 \lesssim t \lesssim 1.2 \times 10^4$ s. Such a flat $L(t)$ curve signifies that particles are \emph{not} exchanging neighbors. The second feature to note in Figure~\ref{fig:srotation}b is the sharp increases in $\logLt$ beginning around $t \approx 1.5 \times 10^4$~s, for all choices of projection axis.  Such steps correspond to a large number of crossings, no matter the chosen orientation, and therefore signify a collective rearrangement of particles (cage-breaking event).  Another such event can be seen in Fig.~\ref{fig:braid} just after $t = 1 \times 10^4$~sec.  Due to these events, the slope of $\logLt$ can be much larger on short timescales. As such, the assumption that $\logLt$ grows linearly in time is \emph{not} valid for short timescales. However, we find $\Sbraid$ is well-defined for systems where  ($\mphi \lesssim 0.79$, $P \lesssim 50$) as $\logLt$ is  linear on the experimental timescale, extensive, and independent of axis of projection.

\subsection{Comparison}

\begin{figure}
\parbox{0.5\textwidth}{\includegraphics[width=0.99\linewidth]{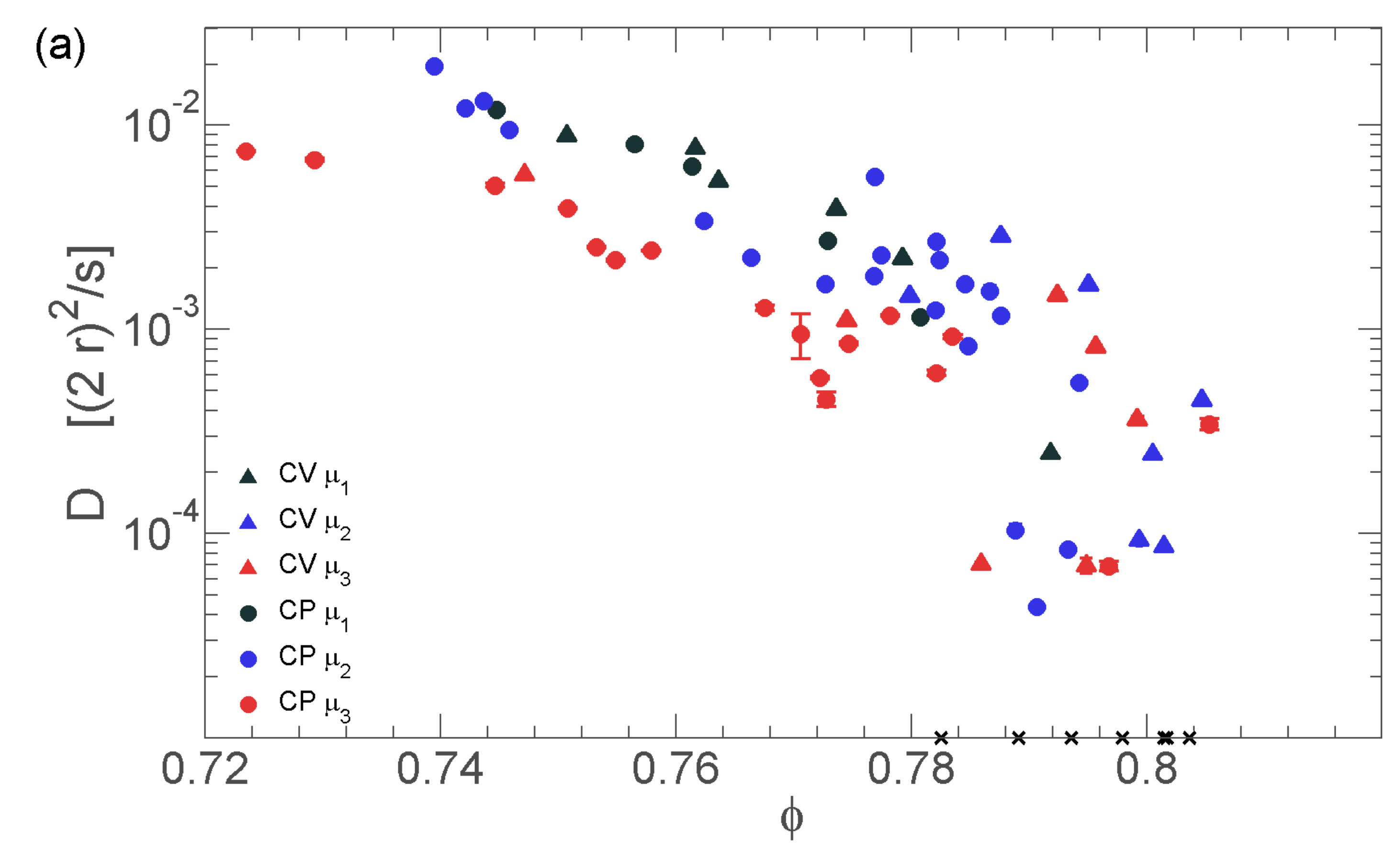}}
\parbox{0.5\textwidth}{\includegraphics[width=0.99\linewidth]{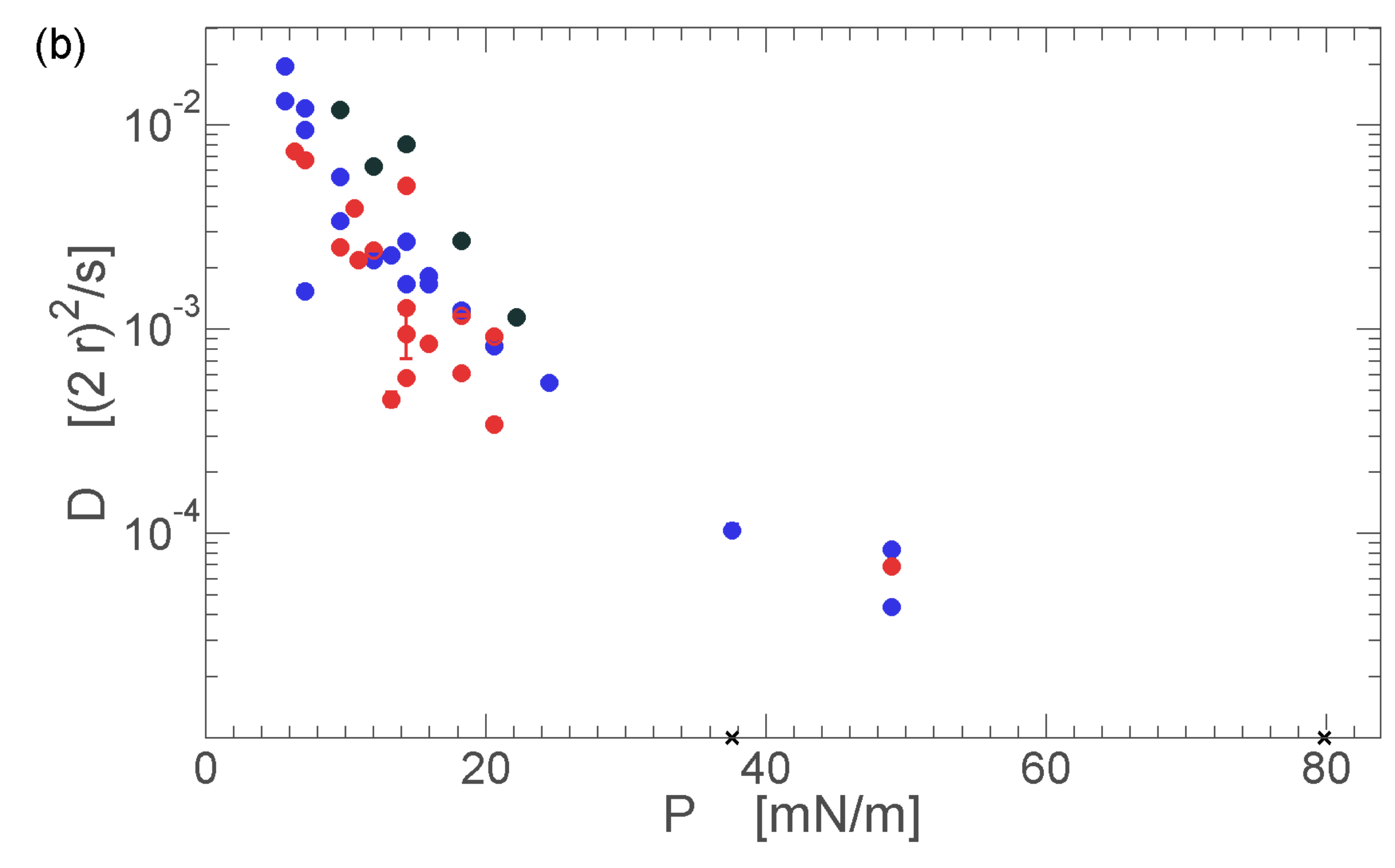}}
\parbox{0.5\textwidth}{\includegraphics[width=0.99\linewidth]{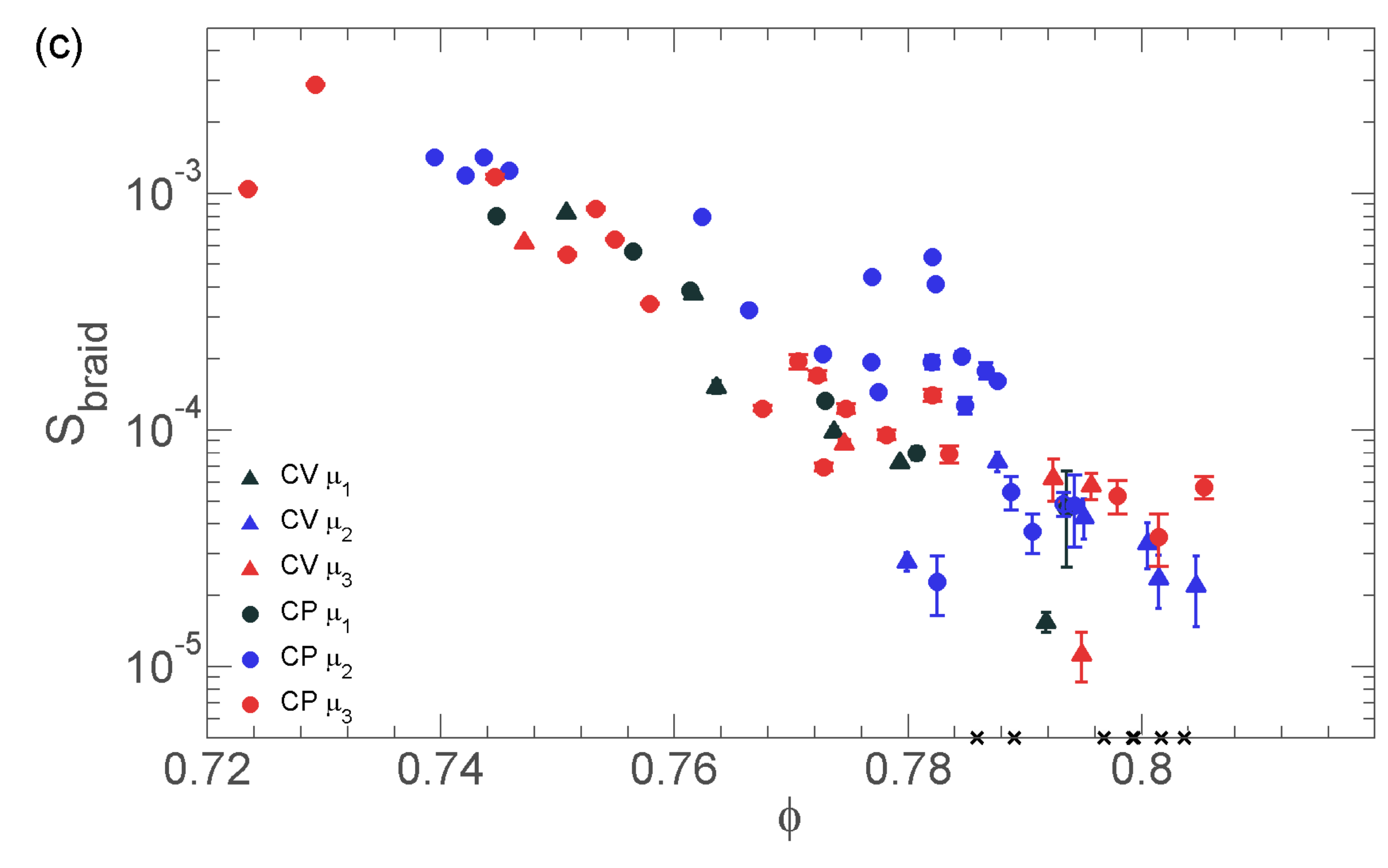}}
\parbox{0.5\textwidth}{\includegraphics[width=0.99\linewidth]{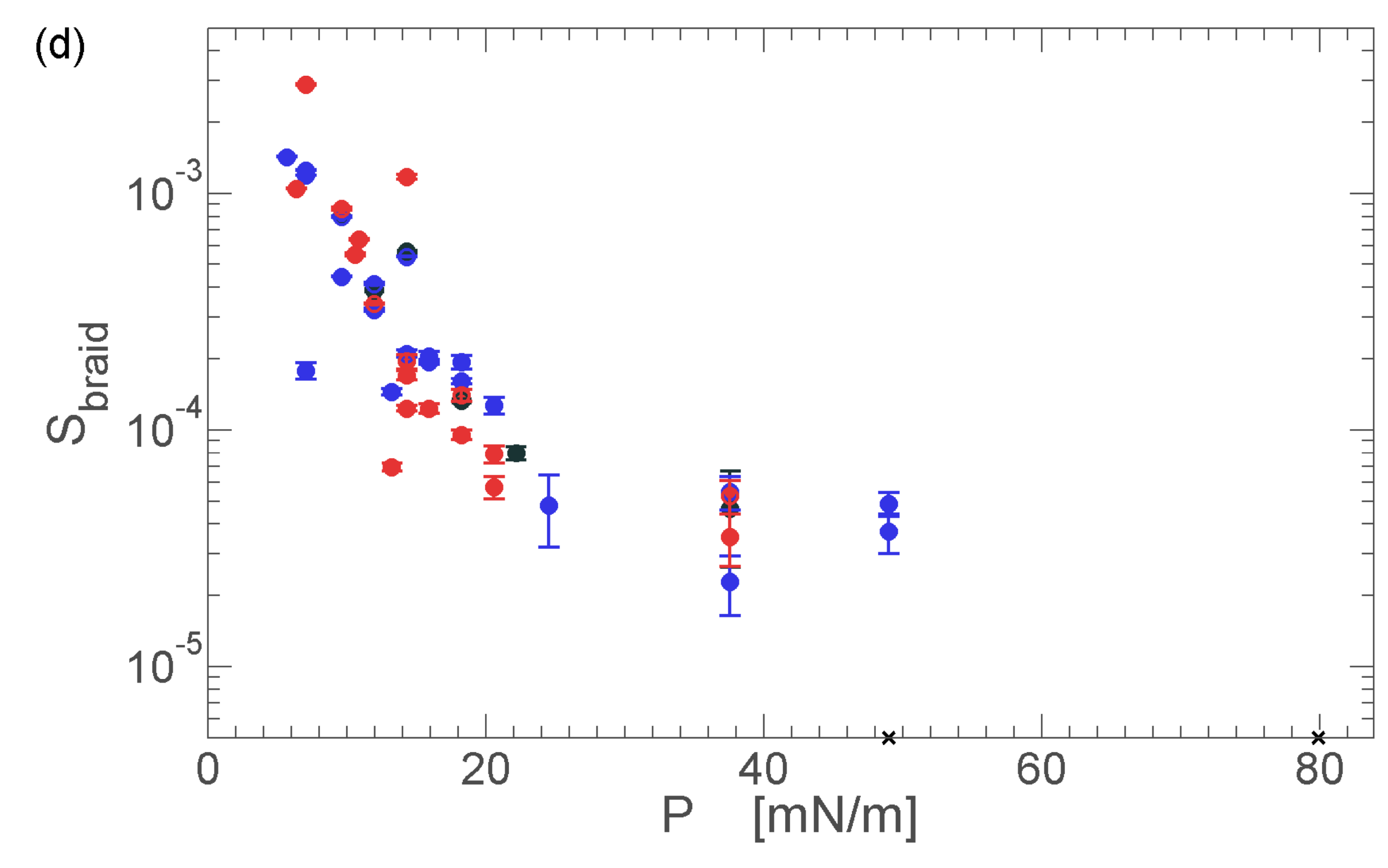}}
\center{\parbox{0.5\textwidth}{\includegraphics[width=0.99\linewidth]{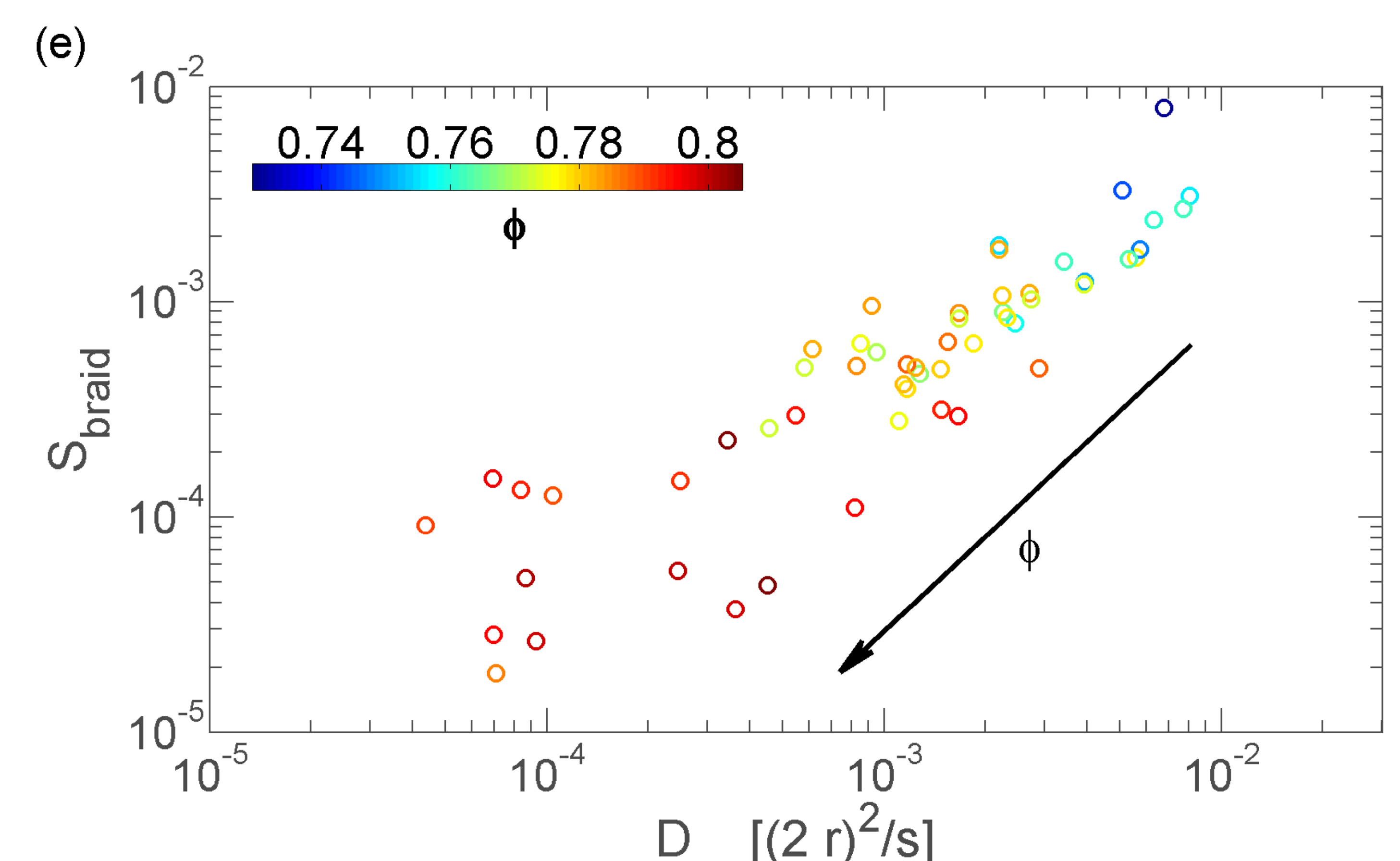}}}
\caption{ Measured values of $D$ as a function of $\mphi$ and $P$ in (a) and (b), and likewise $\Sbraid$ as a function of $\mphi$ and $P$ in (c) and (d), respectfully. Error bars denote the magnitude of the standard error. The black $\times$ on the axis signifies systems for which $D$ or $\Sbraid$ is poorly defined for that particular $\mphi$ or $P$. 
(e) A logarithmic plot of $\Sbraid$ and $D$ for each system, where the arrow denotes increasing $\mphi$. 
}
\label{fig:comparison}
\end{figure}

Based on prior work \cite{Weeks2002, Abate2006, Lechenault2010}, we expect $D$ to decrease as $\mphi$ increases.  Indeed, in Figure~\ref{fig:comparison}ab, $D$ decreases with increasing $\mphi$ and $P$, independent of boundary condition (CP and CV).  Similarly, we find that $\Sbraid$ decreases with increasing $\mphi$ and $P$, as shown in Figure~\ref{fig:comparison}cd, respectively. Error bars represent the average standard error of $\Sbraid$ from the set of 9 projections on axes oriented from $\theta = 0$ to $\pi/2$, where the magnitude of the standard error of $\Sbraid$ is $\approx 10\%$ and does not depend on $\mphi$ and $P$.  We observe little dependence on the inter-particle friction coefficient $\mu$ on $D$ and $\Sbraid$, though qualitatively $D$ decreases with increasing $\mu$.  

As discussed earlier, an experimental measurement of $D$ using $\textnew{\MSD}$ requires a constraint on the value of $\alpha$ to be used to find $D$, which was selected to be $| \alpha-1 | < 0.1$.  For systems with high $P$, this constraint was not met and therefore $D$ is poorly defined, as shown in Figure~\ref{fig:comparison}ab. Similarly, $\Sbraid$ is also poorly-defined for high $P$ systems as $\logLt$ no longer grows linearly within the experimental timescale. For systems with $\mphi \gtrsim 0.79$ and $P \gtrsim  50$~mN/m, neither $D$ nor $\Sbraid$ are useful for measuring the dynamics even with long experimental timescales of $5 \times 10^4$s ($\approx 14$ hr).

\section{Discussion \& Conclusion}

We have investigated the behavior of the global braid entropy $\Sbraid$ associated with the long-time dynamics of a dense granular assembly, and find it to be a well-defined quantity. Notably, $\Sbraid$ was observed to be extensive, making it a promising candidate to play a role in the elaboration of a proper equation of state for dense granular systems. The breakdown in the measurement of $\Sbraid$ coincides with the loss of diffusive dynamics, which is to be expected due to the association of the onset of caging behavior with a loss of ergodicity. The measurement of both $\Sbraid$ and the single-particle diffusion coefficient $D$  provide a quantitative measure of the degree to which slower individual dynamics also results in poorer mixing. In spite of the fact that $D$ is a single-particle measurement, we find that it is a good predictor of the degree of mixing. One benefit of characterizing the average dynamics with $\Sbraid$ instead of $D$ is that no arbitrary fitting parameters are necessary: the method is fully specified by the trajectories themselves.

In addition to providing a measure of the average behavior, the braiding factor $L(t)$ provides a simple way to identify temporal events governing the dynamics (the crossings). In particular, $L(t)$ displays increasing intermittency as the packing density or pressure is increased, with successions of plateaus and jumps indicative of collective neighborhood changes, or structural rearrangements. In fact, this departure from linearity can provide a measure of the chaoticity/ergodicity of the dynamics. The duration required to reach an acceptable degree of linearity, independent of projection axis, is a measurement of a characteristic dynamical timescale of the system. From this point of view, the experimental timescales over which we have investigated our dense granular systems were comparable or larger than this equilibration time, with the exception of the highest $P$ and $\mphi$ systems we tested.  For our experimental timescale, a loss of ergodicity was observed for systems with close proximity to the static jamming transition. 

This investigation represents a first step towards a more systematic and controlled characterization of the braid entropy in two-dimensional granular systems. The braiding factor $L(t)$ may be a means of obtaining system trajectories which could be analyzed with  the recently developed thermodynamics of histories \cite{garrahan2009first}. Quantitative benchmarking of $\Sbraid$ in thermal and glassy systems, together with a complete comparison with the usual dynamical susceptibilities are promising directions for future investigations.

\ack{We are grateful to the National Science Foundation for providing support under grant numbers DMR-0644743 (JGP and KED) and DMS-0806821 (J-LT).}

\section*{References}


\end{document}